\documentclass[runningheads]{llncs}
\usepackage[T1]{fontenc}
\usepackage{graphicx}
\usepackage{marvosym}

\usepackage{multirow}

\begin{document}
\title{A Segmentation Foundation Model for Diverse-type Tumors}
\titlerunning{A Segmentation Foundation Model for Diverse-type Tumors}

\author{Jianhao Xie\inst{1} \and
Ziang Zhang\inst{1} \and
Guibo Luo\inst{1}\textsuperscript{\Letter} \and
Yuesheng Zhu\inst{1}\textsuperscript{\Letter}}

\authorrunning{Jianhao, Ziang et al.}

\institute{School of Electronic and Computer Engineering,Peking University,China 
Correspondences:\email{luogb@pku.edu.cn, zhuys@pku.edu.cn}\\}

\maketitle              
\begin{abstract}
Large pre-trained models with their numerous model parameters and extensive training datasets have shown excellent performance in various tasks. Many publicly available medical image datasets do not have a sufficient amount of data so there are few large-scale models in medical imaging. We propose a large-scale Tumor Segmentation Foundation Model (TSFM) with 1.6 billion parameters using Resblock-backbone and Transformer-bottleneck,which has good transfer ability for downstream tasks. To make TSFM exhibit good performance in tumor segmentation, we make full use of the strong spatial correlation between tumors and organs in the medical image, innovatively fuse 7 tumor datasets and 3 multi-organ datasets to build a 3D medical dataset pool, including 2779 cases with totally 300k medical images, whose size currently exceeds many other single publicly available datasets. TSFM is the  pre-trained model for medical image segmentation, which also can be transferred to multiple downstream tasks for fine-tuning learning. The average performance of our pre-trained model is 2$\%$ higher than that of nnU-Net across various tumor types. In the transfer learning task, TSFM only needs 5$\%$ training epochs of nnU-Net to achieve similar performance and can surpass nnU-Net by 2$\%$ on average with 10$\%$ training epoch. Pre-trained TSFM and its code will be released soon.

\keywords{Medical image segmentation \and Foundation Model\and Transfer Learning.}
\end{abstract}
\section{Introduction}
Medical image segmentation is important in medical image processing~\cite{ref25}. In previous research, deep learning based models like U-Net~\cite{ref1}, nnU-Net~\cite{ref2}, 3D- TransUNet~\cite{ref6} and SwinUNETR~\cite{ref3} have achieved great success. However, many current methods are built for specific tasks and trained on a single dataset, lacking a general model that can perform well on various datasets. When confronted with other tasks, these models need to be re-trained and it is uncertain whether their performances are stable. Take nnU-Net for an example, although it is an outstanding and general framework applicable to various tasks, it still requires re-training from the ground up when applied to a particular task. Conversely, large-scale models may have greater potential compared to other traditional models~\cite{ref1,ref2}. Large-scale models can effectively model complex image features, maintaining high accuracy and generalization performance on multiple datasets. In downstream tasks, they can achieve high accuracy with zero or only a small number of samples for fine-tuning. We introduce a foundational segmentation model designed for diverse tumor types. Previous models~\cite{ref9,ref10} have achieved high accuracy in multi-organ segmentation, but due to the complexity and variability of tumor structures, the segmentation accuracy is still very low. Using diverse tumor imaging data to train large-scale models to achieve general segmentation of multiple tumors can greatly reduce the workload of experts~\cite{ref28}. Currently, large-scale models in the medical field are mainly used for natural language-related medical diagnosis, such as the automatic generation of medical image descriptions~\cite{ref24}.  In the field of medical imaging, STU-Net~\cite{ref5} is a based on the U-Net structure, scalable and transferable medical image segmentation models, and its number of parameters has been expanded to obtain a certain degree of generalization ability. However, in specific fields such as tumor imaging segmentation, large models are relatively scarce~\cite{ref29}. In terms of foundation models for tumor segmentation, several key issues need to be addressed:

\noindent{(1) \textbf{Data challenges in medical imaging}} Acquiring a substantial amount of segmentation data in medical imaging, particularly in tumor segmentation is challenging. Effectively integrating multimodality and multi type data and mitigating model overfitting due to limited data are critical concerns.

\noindent{(2) \textbf{Training of large-scale models}} Large-scale models come with a significant number of parameters. Designing effective attention mechanisms, residual blocks, and training strategies is essential 
to prevent large-scale model collapse and minimize redundancy.

\noindent{(3) \textbf{Application to downstream tasks}}
 There is a desire for large-scale models to excel in downstream tasks. Exploring methods for improved transfer learning to enhance performance in these tasks is of exceptional interest. 

In response to the above issues, we propose TSFM, which utilizes a Resblock-backbone and Transformer-bottleneck architecture. With 1.6 billion parameters, TSFM is capable of learning sufficiently complex features and exhibits excellent generalization ability. The network structure of TSFM allows for transferability to various downstream tasks. To ensure accurate tumor segmentation results and sufficient data, we integrate multiple datasets and create a dataset pool while carefully controlling its composition. Following the integration, we train TSFM on this dataset pool.

Our contribution can be summarized as:

(1)	We propose Tumor Segmentation Foundation Model (TSFM) which has 1.6 billion parameters and a concise and effective network structure.

(2)	We build a 3D medical image dataset pool from multiple public datasets and organize its components specially for TSFM to learn the spatial correlation between tumors and organs comprehensively.

(3)	After training on the dataset pool, TSFM can greatly improve the accuracy of tumor segmentation and quickly transfer to downstream datasets, achieving satisfactory results.

\begin{figure}[t]
\includegraphics[height=8cm,width=\textwidth]{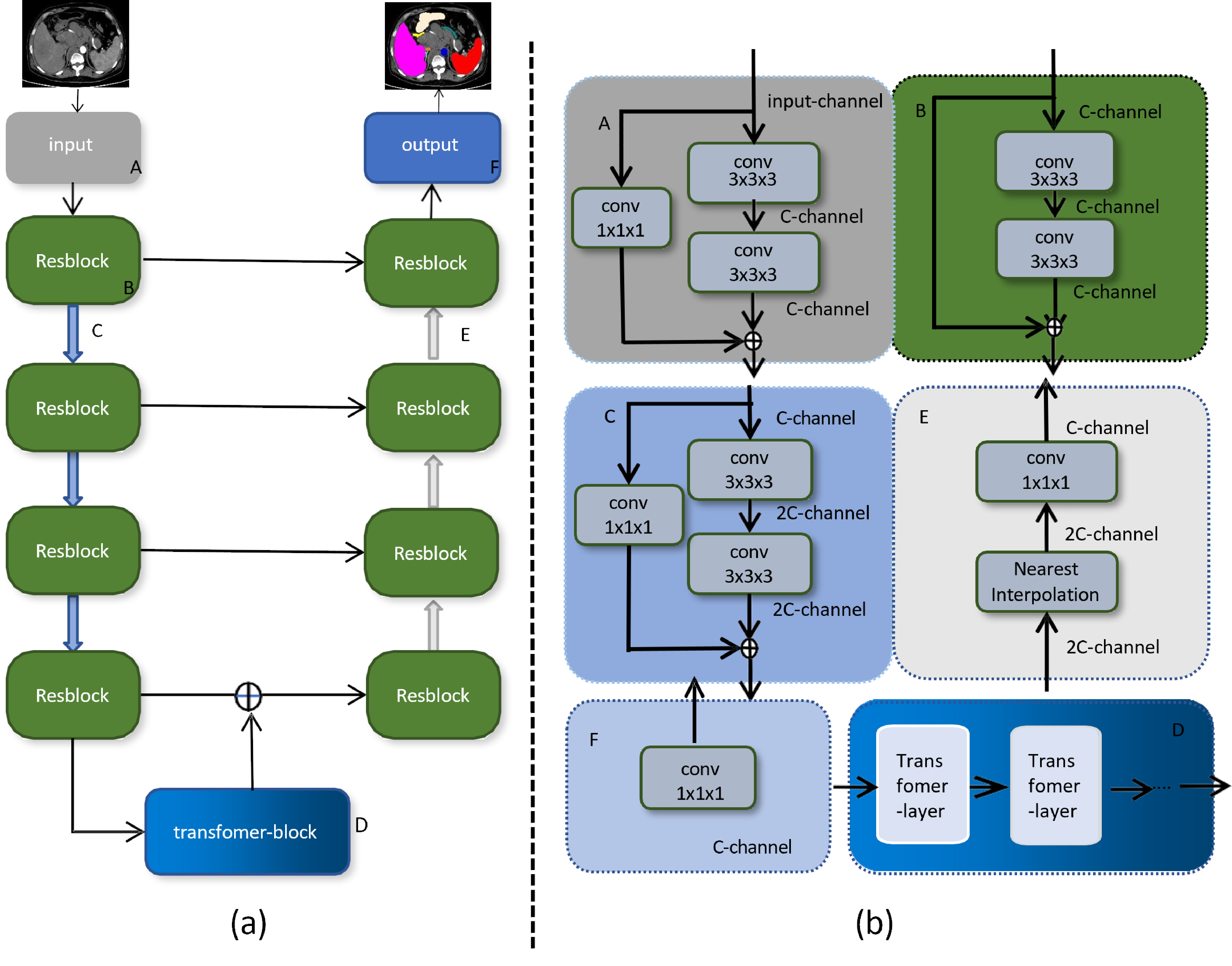}
\caption{Network structure. Part (a) shows the overall structure of TSFM. Part (b) shows the internal structure of the specific block. TSFM consists of six main blocks: A, B, C, D, E, and F.} \label{fig1}
\end{figure}

\section{Method}
Our main works are as follows: we utilize the effective data pre-processing mechanism of nnU-Net framework before training our foundation model. Then we construct the dataset pool with the consideration of the space correlation between multi-organs and tumors which will be used for TSFM pre-training. After model pre-training, we transfer it to several downstream datasets and achieved excellent results with only a short training time. We demonstrate the versatility of the pre-trained model and its strong transferability in tumor segmentation.
\subsection{Dataset Pool}
Prior to initiating model training, it is imperative to address the inadequacy of available data. To tackle this challenge, we adopt a dataset integration method akin to the recent work MultiTalent~\cite{ref7}. This method seamlessly combines multiple datasets, effectively mitigating the issue of distinct label IDs for the same target across different datasets. The method employ the binary Binary Cross-entropy loss (BCE) and a modified Dice loss for each class over all images in a batch. And to compensate for the varying number of training images in each dataset, we choose a sampling probability per case that is inversely proportional to $\sqrt{n}$, where $n$ is the number of training cases in the corresponding source dataset. In our approach, we integrate ten datasets to construct dataset pool, comprising seven tumor datasets (Liver, Lung, HepaticVessel, Colon, Pancreas from MSD~\cite{ref17}, Breast cancer~\cite{ref27}, KiTS2021~\cite{ref21}) and three multi-organ datasets (AMOS22~\cite{ref16}, BTCV~\cite{ref18}, Totalsegmentor-part1~\cite{ref15}). As detailed in Section 1, we create a dataset pool to address the scarcity of tumor data and enhance the tumor segmentation capabilities of TSFM. Directly fusing datasets will cause a challenge of inconsistent labels, where in the labels for the same type of target in different datasets vary. To overcome this, we establish a label remapping table to correlate the labels of the original datasets with the fused dataset pool. This ensures that each target has a unique label during training. Noteworthy is the innovative fusion of multi-organ datasets and tumor datasets, distinguishing our approach from MultiTalent.  
\subsection{Network Structure}
Our network architecture preserves the U-shaped structure and skip connections, integrating a Resblock-backbone and Transformer-bottleneck. This design facilitates the extraction of finer-grained features compared to pure convolutional networks while addressing concerns regarding excessive memory consumption. Fig 1(a) illustrates the network structure, while detailed information about each block is provided in Fig 1(b).

\noindent{\textbf{Resblock-Backbone:} Same as B-Block in Fig 1(b). This block consists of two normal Conv-IN-LeakyReLU~\cite{ref9,ref11} layers and one residual connection. In the residual connection, we choose pre-activation~\cite{ref19} to ensure that the input parameters are not all positive. In the two normal convolution operations, we use 3x3x3 convolution kernel with padding and stride size of 1. In each resblock, we use multiple stacked basic residual blocks.}

\noindent{\textbf{Down-sampleblock:} Down-sampleblock is based on the resblock with the output channels being twice as many as the input channels. The stride and padding of the convolution are  modified to make the vector size half of the input size.The specific structure is shown in Fig 1 (b) C-Blcok.}

\noindent{\textbf{Upsampleblock:} We use an interpolation operation based on 1x1x1 to do upconvolution, which doubles the size of the channels. Then, we attach a 1x1x1 convolution layer to reduce the number of channels to half of the original. The specific structure is shown in Fig 1 (b) E-Blcok.}

\noindent{\textbf{Transformer-Bottleneck:} Considering that transformer requires a large amount of CUDA memory, we choose to add it to the lowest layer of our U-shaped network. We use the ViT-base~\cite{ref20} version, which includes a 12-layer transformer encoder. The specific structure is shown in Fig 1 (b) D-Blcok.}


\subsection{Transfer Learning}
Transfer learning in deep learning typically involves leveraging pre-trained models to initialize new models, thereby transferring learned features to related tasks. Our proposed TSFM demonstrates strong transferability due to the following design considerations:
(1) The TSFM architecture combines CNN with transformer, enabling it to maintain learned weights from the CNN component even when dealing with downstream images of varying sizes.  This ensures that most of the learned weights on CNN are preserved, with adjustments made only to the transformer-bottleneck part.
(2) The concise network structure of TSFM reduces the risk of overfitting, enhances transferability, and yields improved transfer performance.
(3) The dataset pool used for training TSFM encompasses diverse variations, multiple tumor types, and a sufficient number of datasets. This rich dataset pool effectively supports TSFM in extracting foundational features from medical images, making it suitable for various other medical segmentation tasks.

\begin{table}
\centering
\caption{Comparison of tumor dataset performance.TSFM and nnU-Net-datasetpool are trained 1,000 epochs on 7 tumor datasets. Others are trained in different datasets to train 1,000 epochs respectively. The scores in the table are Dice values.}\label{tab 1}
\setlength{\tabcolsep}{0.5mm}{
\begin{tabular}{|c|cc|c|cc|cc|} 
\hline
                    & \multicolumn{2}{c|}{Liver}        & Lung            & \multicolumn{2}{c|}{Pancreas}     & \multicolumn{2}{c|}{HepaticVessel} \\ \hline
                    & liver           & tumor           & tumor           & pancreas        & tumor           & vessel           & tumor           \\ \hline
nnU-Net             & 0.9621          & 0.5781          & 0.6216          & \textbf{0.8092} & 0.5791          & 0.6400           & 0.7360          \\
SwinUNETR-base      & 0.9297          & 0.4927          & 0.5257          & 0.7471          & 0.5121          & 0.5884           & 0.6142          \\
nnU-Net-datasetpool & 0.9579          & 0.5722          & 0.5884          & 0.7938          & 0.5643          & 0.6377           & 0.7315          \\
SAM-Med3D &
  \multicolumn{1}{l}{0.9235} &
  \multicolumn{1}{l|}{0.5387} &
  \multicolumn{1}{l|}{0.4492} &
  0.5768 &
  \multicolumn{1}{l|}{0.3144} &
  \multicolumn{1}{l}{0.1564} &
  \multicolumn{1}{l|}{0.5685} \\
 
Ours                & \textbf{0.9619} & \textbf{0.6209} & \textbf{0.6391} & 0.8071          & \textbf{0.6535} & \textbf{0.6426}  & \textbf{0.7462} \\ \hline
\end{tabular}}
\setlength{\tabcolsep}{2.18mm}{
\begin{tabular}{|c|c|cc|c|cl|}
\hline
 & Colon & \multicolumn{2}{c|}{KiTS2021} & Breast & \multicolumn{2}{c|}{\multirow{2}{*}{\begin{tabular}[c]{@{}c@{}}mean Dice\\ for tumor\end{tabular}}} \\ \cline{1-5}
                           & tumor           & kidney          & tumor           & tumor           & \multicolumn{2}{c|}{}   \\ \hline
nnU-Net                    & 0.5271          & \textbf{0.9684} & \textbf{0.8695} & 0.7451          & \multicolumn{2}{c|}{0.6652} \\
SwinUNETR-base             & 0.4880          & 0.9135          & 0.5788          & 0.6446          & \multicolumn{2}{c|}{0.5508} \\
nnU-Net-datasetpool        & 0.5815          & 0.9618          & 0.8328          & 0.7261          & \multicolumn{2}{c|}{0.6566} \\
SAM-Med3D                  & 0.4554          & 0.6835          & 0.4426          & 0.7524          & \multicolumn{2}{c|}{0.5030} \\
Ours                       & \textbf{0.6620} & 0.9650          & 0.8608          & \textbf{0.7923} & \multicolumn{2}{c|}{\textbf{0.7106}} \\ \hline
\end{tabular}}
\end{table}
\begin{figure}[t]
\includegraphics[height=7cm,width=\textwidth]{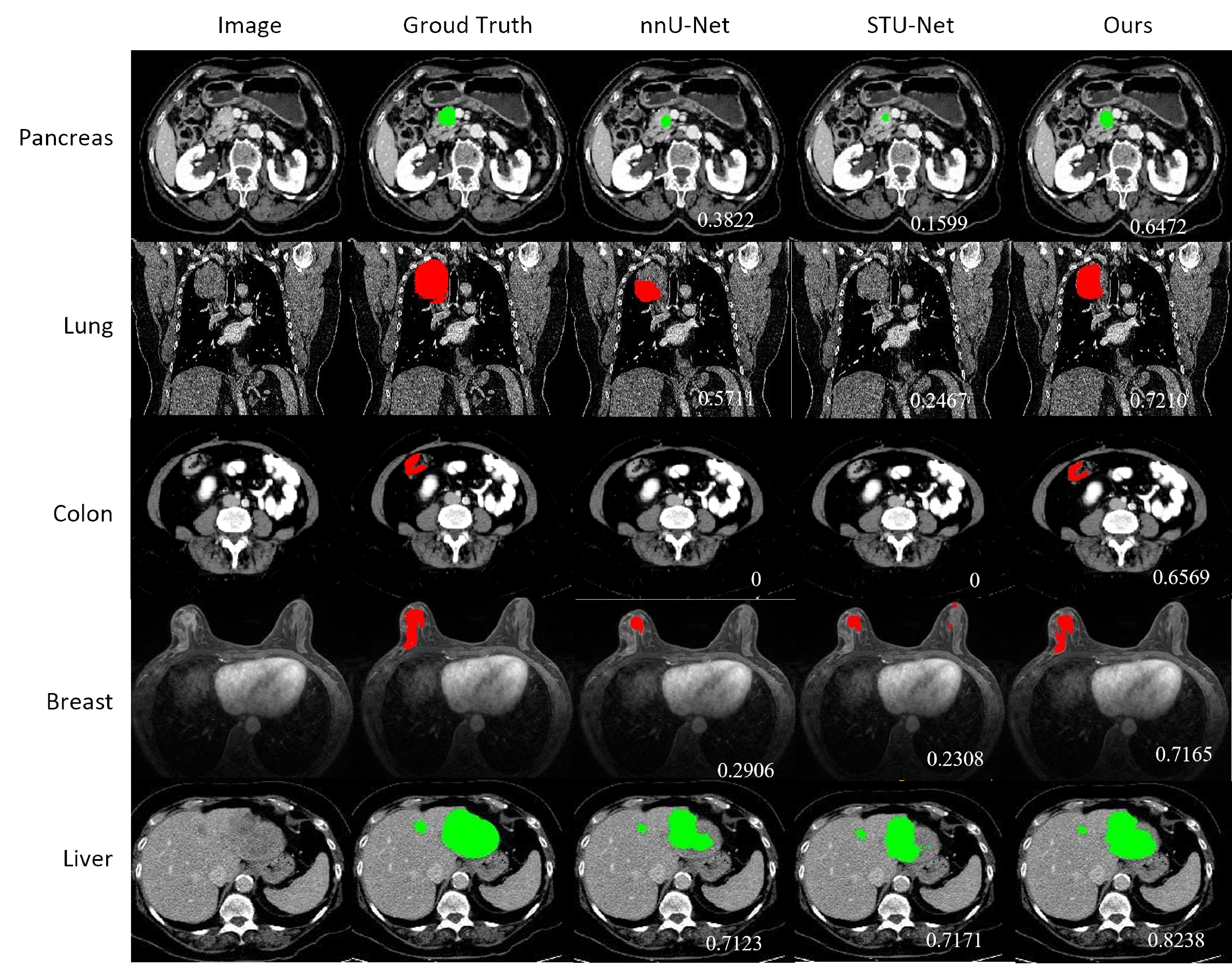}
\caption{Segmentation Results.} \label{fig2}
\end{figure}

\section{Experiments}
\subsection{Details of experiment}
The environment of the experiments is all based on Python 3.9.18, CentOS 7, Pytorch 2.1.0+cu121 and nnU-Net v1.

We utilizes preprocessing, postprocessing, and similar operations to those in the nnU-Net v1 framework to process the data. The optimizer for the model is selected as SGD optimizer with Nestrov momentum of 0.99 and a weight decay of $1e^{-3}$. The batch size is fixed to 2 and each epoch contains 250 iterations, with a total of 1000 epochs, an initial learning rate of 0.001, and a patch size of 112x160x192. In the transfer experiment, all parameters will be set according to the default settings of nnU-Net. 

\subsection{Comparison on each dataset of dataset pool}
The comparison in this section involves TSFM, nnU-Net, Swin UNTER, and SAM-Med3D~\cite{ref12,ref13,ref30}, with separate Dice scores provided for each type of data. nnU-Net and Swin UNTER are trained using their default parameters from the original official code implementation on 10 datasets separately for 1000 epochs. The experimental results of SAM-Med3D are obtained using its pre-trained model with original parameters. Detailed data is presented in Table 1, which lists all tumor datasets. Due to space constraints, the results for multi-organ datasets will be provided in the supplementary materials. As depicted in Table 1, our model demonstrates strong performance on tumor datasets, surpassing nnU-Net in most cases, with significant improvements observed. These results underscore the effectiveness of our model, achieving an average improvement of 3$\%$ over nnU-Net in tumor segmentation and comparable performance in multi-organ segmentation. For specific results, please refer to Fig. 2.

\begin{table}
\caption{Comparison of transfer learning on BraTS2020, KiTS2019 and Adbomen-1K. After 100 epochs of transfer training, TSFM can achieve an average performance improvement of about 3\% compared to nnU-Net trained for 1000 epochs.  }\label{table3}
\setlength{\tabcolsep}{4.00mm}{
\begin{tabular}{|p{3.8cm}|p{1cm}|p{1cm}||p{1cm}|p{1cm}|}
\hline
\textbf{BraTS2020/tumor       } & \multicolumn{1}{c|}{10epoch} & \multicolumn{1}{c|}{20epoch} & \multicolumn{1}{c|}{50epoch} & 100epoch        \\ \hline
Ours-pretrain           & \multicolumn{1}{c|}{0.5158}  & \multicolumn{1}{c|}{0.5355}  & \multicolumn{1}{c|}{0.5481}  & \textbf{0.5571} \\
Ours-nopretrain   & \multicolumn{1}{c|}{0.1023} & \multicolumn{1}{c|}{0.2056} & \multicolumn{1}{c|}{0.2789} & 0.3763 \\
STU-Net-pretrain  & \multicolumn{1}{c|}{0.5162} & \multicolumn{1}{c|}{0.5362} & \multicolumn{1}{c|}{0.5475} & 0.5548\\ \cline{2-5} 
nnU-Net-1000epoch & \multicolumn{4}{c|}{0.5553}                                                                      \\ \hline
\end{tabular}}
\setlength{\tabcolsep}{4.00mm}{
\begin{tabular}{|p{3.8cm}|p{1cm}|p{1cm}||p{1cm}|p{1cm}|}
\hline
\textbf{KiTS2019/tumor            } & \multicolumn{1}{c|}{10epoch} & \multicolumn{1}{c|}{20epoch} & \multicolumn{1}{c|}{50epoch} & 100epoch        \\ \hline
Ours-pretrain           & \multicolumn{1}{c|}{0.6723}  & \multicolumn{1}{c|}{0.7953}  & \multicolumn{1}{c|}{0.8045}  & \textbf{0.8503} \\
Ours-nopretrain   & \multicolumn{1}{c|}{0.1306} & \multicolumn{1}{c|}{0.2221} & \multicolumn{1}{c|}{0.5589} & 0.7413 \\
STU-Net-pretrain  & \multicolumn{1}{c|}{0.6146} & \multicolumn{1}{c|}{0.7182} & \multicolumn{1}{c|}{0.8003} & 0.8036 \\ \cline{2-5} 
nnU-Net-1000epoch & \multicolumn{4}{c|}{0.8454}                                                                      \\ \hline
\end{tabular}}
\setlength{\tabcolsep}{4.0mm}{
\begin{tabular}{|p{3.8cm}|p{1cm}|p{1cm}||p{1cm}|p{1cm}|}
\hline
\textbf{Abdomen/multi-organ} & \multicolumn{1}{c|}{10epoch} & \multicolumn{1}{c|}{20epoch} & \multicolumn{1}{c|}{50epoch} & 100epoch        \\ \hline
Ours-pretrain                & \multicolumn{1}{c|}{0.9309}  & \multicolumn{1}{c|}{0.9355}  & \multicolumn{1}{c|}{0.9402}  & \textbf{0.9437} \\
Ours-nopretrain   & \multicolumn{1}{c|}{0.6946} & \multicolumn{1}{c|}{0.7963} & \multicolumn{1}{c|}{0.8686} & 0.9025 \\
STU-Net-pretrain  & \multicolumn{1}{c|}{0.9264} & \multicolumn{1}{c|}{0.9395} & \multicolumn{1}{c|}{0.9416} & 0.9426 \\ \cline{2-5} 
nnU-Net-1000epoch & \multicolumn{4}{c|}{0.9337}                                                                      \\ \hline
\end{tabular}}
\end{table}

\subsection{Transfer learning comparison}
To evaluate the transferability of TSFM, we deploy it in other downstream tasks, including BraTS2020, KiTS2019~\cite{ref22}  and Adbomen-1K~\cite{ref14}, which are not included in the original dataset pool. We design experiments with 4 different training epochs (10, 20, 50, 100). Regarding the loading model, we use the best model stored in pre-training for downstream tasks. We conduct training both with and without loading the pre-trained model weights across these epochs. Additionally, we compare our approach with STU-Net~\cite{ref5}, known for its strong transferability in medical image segmentation tasks, using its huge version for the same transfer process. Finally, we compare the results with nnU-Net trained for 1000 epochs.

Among the three datasets, BraTS2020 and KiTS2019 are tumor datasets comprising 368 and 210 cases, respectively. In the transfer learning experiments, our focus is on comparing the tumor segmentation performance. Abdomen-1K, on the other hand, is a multi-organ dataset consisting of 50 cases, and we compare the average segmentation scores across all organs.
The detailed results are shown in Table 2. From the table, it can be seen that the pre-trained TSFM has excellent transfer learning ability. TSFM shows good performance after fine-tuning at 100 epochs, surpassing the nnU-Net results trained at 1000 epochs and also surpassing the STU-Net fine-tuning results at 100 epochs. After only training for 50 epochs, the performance is also close to the result of nnU-Net. These results indicate that TSFM can migrate to downstream datasets with only 10$\%$ of the trained epoch, and achieve good performance. With only 5$\%$ of the trained epoch, we can achieve a performance close to nnU-Net. The remarkable performance of our model in transfer learning can be attributed to the meticulous design of the model structure, the abundance of parameters, and the extensive dataset, underscoring the robust transferability of TSFM. Additionally, the enhancement of the network structure through the incorporation of a transformer component proves instrumental in optimizing tumor segmentation.

\begin{table}
\caption{Comparison of dataset ablation results. We
use two different single datasets to pre-train our model and then
performed 100 epochs of transfer training}\label{table4}
\setlength{\tabcolsep}{7.29mm}{
\begin{tabular}{|c|c|c|c|}
\hline
               & BraTS2020       & KiTS2019        & Abdomen         \\ \hline
Ours-pretrain  & \textbf{0.5571} & \textbf{0.8503} & \textbf{0.9437} \\
Ours(AMOS22)   & 0.5446         & 0.8329          & 0.9397          \\
Ours(Pancreas) & 0.5423         & 0.8219          & 0.9245          \\ \hline
\end{tabular}}
\end{table}

\subsection{Ablation experiment}
Since the primary contributions of our work lie in the model architecture and dataset pool construction, ablation experiments are conducted on these two aspects.

\textbf{Model ablation} To demonstrate the effectiveness of our model, we compare it with nnU-Net trained on the dataset pool (the same dataset used for TSFM’s training) for 1000 epochs. Comparing the results of nnU-Net and TSFM, it becomes evident that the TSFM network structure outperforms nnU-Net. Detailed performance metrics are presented in Table 1, illustrating the superior performance of our method when compared to nnU-Net trained on the dataset pool. Our model's advantage lies in its ample number of parameters, enabling it to adeptly capture diverse characteristics within larger datasets, thereby yielding superior results. In contrast, nnU-Net, constrained by a relatively limited parameter count, struggles to comprehensively learn each feature within extensive datasets. This discrepancy significantly contributes to the superior performance of our model in ablation experiments.

\textbf{Dataset pool ablation} To validate the effectiveness of the dataset pool, we train and transfer our model on each dataset within the pool. We compare the results of TSFM trained on the dataset pool with those trained on individual datasets to demonstrate the promoting effect of the dataset pool on model performance. We select two representative datasets for our experiments: a tumor dataset (Pancreas) and a multi-organ dataset (AMOS22). Training our model on each dataset for 1000 epochs and then performing downstream transfer learning on the same dataset, we compare the results with those obtained in Section 3.3. As depicted in Table 3, our model exhibits significantly better transferability when trained with the dataset pool compared to training without it. The ablation experiment on the dataset underscores the importance of a sufficient dataset in the pre-training process of the cornerstone model. When the dataset is inadequate, the model's performance may be compromised, highlighting the crucial role of a comprehensive dataset in achieving optimal results.

\section{Conclusions}
We build a dataset pool comprising 2779 cases, totaling 300k medical images, and design a network architecture for TSFM by using Resblock-backbone and Transformer-bottleneck with 1.6 billion parameters. TSFM outperforms previous methods in the segmentation of diverse tumor types and demonstrates competitive performance in multi-organ segmentation. Through conducting downstream transfer learning, our model only needs 10$\%$ training epochs of nnU-Net to surpass its performance, revealing that our model exhibits stronger transfer ability in tumor segmentation. 


%
%
%
%

\end{document}